\documentclass[aoas,nameyear,seceqn,dvips]{arximspdf}
\usepackage{multirow}
\usepackage{graphicx}

\doi{10.1214/10-AOAS366}
\volume{4}
\issue{4}
\pubyear{2010}
\firstpage{2181}
\lastpage{2202}



\begin{document}
\begin{frontmatter}

\title{Modeling heterogeneity in ranked responses by nonparametric maximum likelihood:
How do Europeans get their scientific knowledge?}
\runtitle{Heterogeneity in ranked responses}

\begin{aug}
\author[A]{\fnms{Brian} \snm{Francis}\ead[label=e1]{B.Francis@Lancaster.ac.uk}\corref{}},
\author[B]{\fnms{Regina} \snm{Dittrich}\ead[label=e2]{Regina.Dittrich@wu.ac.at}}
\and
\author[B]{\fnms{Reinhold} \snm{Hatzinger}\ead[label=e3]{Reinhold.Hatzinger@wu.ac.at}}

\runauthor{B. Francis, R. Dittrich and R. Hatzinger}

\affiliation{Lancaster University, Vienna University of Economics and Business and\\ Vienna University of Economics and Business}
\address[A]{B. Francis\\
Department of Mathematics\\
\quad and Statistics\\
Fylde College\\
Lancaster University\\
Lancaster LA1 4YF\\
UK\\
\printead{e1}} 
\address[B]{R. Dittrich\\
R. Hatzinger\\
Department of Statistics and Mathematics\\
Vienna University of Economics and Business\\
Augasse 2-6\\
A-1090 Wien\\
Austria\\
\printead{e2}\\
\phantom{E-mail:\ }\printead*{e3}}
\end{aug}

\received{\smonth{8} \syear{2009}}
\revised{\smonth{6} \syear{2010}}

\begin{abstract}
This paper is motivated by a Eurobarometer survey on science
knowledge. As part of the survey, respondents were asked to rank
sources of science information in order of importance. The official
statistical analysis of these data however failed to use the
complete ranking information. We instead propose a method which
treats ranked data as a set of paired comparisons which places the
problem in the standard framework of generalized linear models and
also allows respondent covariates to be incorporated.

An extension  is proposed to allow for heterogeneity in the ranked responses.
The resulting model uses a nonparametric formulation of the random effects structure, fitted using the EM algorithm.
Each mass point is multivalued, with a parameter for each item.
The resultant model is equivalent to a covariate latent class model, where the latent class profiles
are provided by the mass point components and the covariates act on the class profiles.
This provides an alternative interpretation of the fitted model. The approach is also suitable for paired comparison data.
\end{abstract}



\end{frontmatter}

\section{Introduction}\label{se1}
Ranked  data commonly arise in many substantive areas such as
psychology, social research and marketing research when the interest is
focused on the relative ordering of  various items, options, stimuli or
objects. A typical aim of such studies is to estimate the mean or
average ordering of a set of items, and to investigate how this
ordering changes with respondent characteristics. This paper focuses on
the analysis of a survey question from a special Eurobarometer survey
on science knowledge, which asked respondents to rank six sources of
science information in order of importance.

Eurobarometer public opinion surveys have been carried out in all
member states of the European Union since 1973. Eurobarometer 55.2
 was a special survey collected in 2001 and designed to elicit information
on European experience and perception of science and technology. 17
countries in total were surveyed---with Northern Ireland, Great
Britain, East Germany and West Germany being treated as separate
countries for the purposes of the survey. Within each country a
multistage sampling scheme was used. Primary sampling units (PSUs) were
randomly selected with probability based on population size after
stratification by administrative region and by the degree of
urbanization. Within each PSU, a cluster of addresses was sampled, and
random route methods were used to select households. Finally, a
respondent was selected at random from within each household. Face to
face interviewing was used to elicit responses.

Our question of interest in this paper is given in Figure \ref{aimssw}.
\begin{figure}
\begin {sffamily}
\begin{small}
\begin{center}
\begin{tabular}[t]{@{}l@{}}
\hline
\bfseries{Eurobarometer 55.2 May--June 2001 Question 5.}\\
Here are some sources of information about scientific developments. \\
Please rank them from 1 to 6 in terms of their importance to you \\
(1 being the most important and 6 the least important)\\
\begin{tabular}{@{}lll@{}}
    (a) & Television &  .....\\
    (b) & Radio        &  .....\\
    (c) & Newspapers and magazines &         .....\\
    (d) & Scientific magazines  &          .....\\
    (e) & The internet &            .....\\
    (f) & School/University &  .....\\
\end{tabular}
\\\hline
\end{tabular}
\end{center}
\end{small}
\end{sffamily}
\caption{The 'Sources of science information' question.} \label{aimssw}
\end{figure}
The survey report [\citet{christ01}] describes how this question was
analyzed. Only the first two rank positions were examined, and the
percentage of times a source was mentioned in either the first or
second position was reported. This was presented as given in Table
\ref{percent1}.

\begin{table}[b]
\caption{Respondents mentioning source of information in first or
second position}\label{percent1}
\begin{tabular}{@{}lccccc@{}}
\hline
$\bolds{a}$&$\bolds{b}$&$\bolds{c}$&$\bolds{d}$&$\bolds{e}$&$\bolds{f}$\\
\textbf{Television}&\textbf{Radio}&\textbf{Press}&\textbf{Scientific}&\textbf{Internet}&\textbf{School and}\\
&&&\textbf{magazines}&&\textbf{university}\\
$\bolds{(\mathit{TV})}$&$\bolds{(\mathit{Radio})}$&$\bolds{(\mathit{Press})}$&$\bolds{(\mathit{SciM})}$&$\bolds{(\mathit{WWW})}$&$\bolds{(\mathit{Edu})}$\\
\hline
60.3\%&27.3\%&37.0\%&20.1\%&16.7\%&20.3\%\\
\hline
\end{tabular}
\end{table}

This method of analysis, however, does not use the respondent's last
four ranked positions, and also does not distinguish in importance
between the first and second ranked position. Thus, information is
wasted and other issues such as the influence of covariates and
respondent heterogeneity are not considered.

We proceed by examining current approaches to ranked data in Section \ref{se2},
before describing our modeling approach in Sections \ref{se3}--\ref{se5}. This
approach combines the modeling of ranked data patterns through the
Bradley--Terry model, We parameterize the items through a set of worth
parameters which sum to 1, and which we allow to depend on covariates.
The model also incorporates discrete or nonparametric (mass-point)
random effects to account for heterogeneity. This model can also be
thought of as a mixture or latent class model on the ranks. Algorithmic
and computational issues are discussed in Section \ref{sec:5}, and the results of
the new analysis on the Eurobarometer question above are discussed in
Section \ref{se7}. The paper finishes with a discussion of the methodology.

\section{Existing approaches to ranked data}\label{se2}
Three simple approaches to analyzing ranked data are common in the
literature. The crudest method is simply to analyze only the first
ranked response, but this wastes information by not using the other
ranks. Another approach is to assume that the rankings are from a
continuous scale, and to analyze mean ranks, perhaps invalidly assuming
normality. A third approach, used by sensory perception researchers,
uses the nonparametric Friedman two-way analysis of variance. This
test, however, simply examines the null hypothesis that the median
ranks for all items are equal, and does not consider any differences in
ranking between respondents [\citet{shes07}]. Moreover, if the Friedman
test rejects the null hypothesis, no quantitative interpretation, such
as  the odds of preferring one item over another, is provided.

All of these simple approaches fail both to consider the underlying
psychological mechanism for ranking, and to formulate correct
statistical models for this mechanism. In contrast, the approach taken
in this paper is statistically more rigorous, and involves modeling the
observed ranks by assuming that they are generated through an
underlying choice or preference model.

There are also a variety of modeling approaches to ranked data. One
common approach assumes that the respondent carries out the ranking by
first choosing the most preferred item, and then the next preferred,
and so on. This has led to the choice set explosion model of
\citet{chap82} and the multistage model of \citet{flig88}. For example,
a series of papers by Gormley and Murphy [\citeauthor{gorm08a} (\citeyear{gorm08b,gorm08a})] have
suggested modeling ranks through the Plackett--Luce and Benter models
and have illustrated the methodology using Irish electoral data.
However, more generally, the choice set approach has the disadvantage
of inconsistency: models which assume instead that respondents first
choose the least preferred, then the next least preferred, and so on
lead to different conclusions and estimates of worths.

Other modeling approaches have assumed an underlying distance metric on
the ranks---thus, \citet{busse07} assumed that differences between
ranks can be measured through the Kendall distance, which measures the
number of adjacent transpositions needed to transform one rank into
another. \citet{Delia05} suggested that a two-component mixture of a
shifted binomial  and a uniform distribution be used to model the rank
of an specific item.

In this paper we assume that a ranking of items is produced by the
respondent making a set of consistent paired comparison experiments,
comparing each item mentally with each of the others, until a
consistent ranking is obtained.

\citet{flig93} described suitable probability models for ranked data
such as the Babington Smith model, where the probability for rankings
are defined via parameters for paired comparisons. The usual model for
paired comparisons [\citet{brad52}] was extended to ranked data by
\citet{mall57} (the Mallows--Bradley--Terry model).

\citeauthor{crit91} (\citeyear{crit91,crit93}) showed that the Mallows--Bradley--Terry model is a
Generalized linear model (GLM) and extended the model by introducing
item-specific variables. We adopt this approach in this paper,
extending it by the addition of respondent covariates and random
effects structures.

Ranked responses will vary between respondents. While measured
covariates can be taken into account [\citet{ditt00}; \citet{francis02}], there
are likely to be other unmeasured or unmeasurable characteristics of
the respondents which will also affect the response. This will give
rise to heterogeneity in the data which need to be taken into account.
One approach is to use a mixing distribution approach. \citet{lanc83}
considered random effects models for paired comparison data and fitted
a beta-binomial distribution. \citet{matt95} later extended the model
to involve ties and used a Dirichlet mixing distribution;
\citet{bock01a} fitted a binomial-Normal distribution.

In this paper we use a random effects approach, but adopt a discrete
nonparametric mass point distribution rather than a continuous mixing
distribution. The use of a discrete distribution both avoids  the
considerable computational complexity of multiple integrals in the
continuous case, and also avoids the need to specify a specific
distribution which may by inappropriate. Heterogeneity in effect is
modeled through the incorporation of a missing latent factor
representing group membership. If there are no respondent covariates,
then the approach reduces to a latent class model [\citet{forma1992}].
While \citet{bock01b}, \citet{croon89} and \citet{gorm08b} have
considered the use of latent class models for ranked data, they take a
choice-based rather than a paired comparison approach.

\section{Ranked data and paired comparisons}\label{se3}

The ranking of items  can be described either by a \textit{rank vector}
(which gives the ranks of the items) or by an \textit{order vector}
(which gives the items in rank order).

Paired comparisons have much in common with ranking tasks. In a paired
comparison task the respondents are asked to choose the preferred item
in each pair of items. The number of pairs for a set of $J$ items is
given by $J \choose 2$. In general, the observed paired comparison
response for two items $i$ and $j$ can be coded as
\[
y_{ij}=\cases{
1&\quad\mbox{if item $i$ is preferred to item $j$ $(i \succ j)$},\cr
-1&\quad\mbox{if item $j$ is preferred to item $i$ $(j \succ i)$}.
}
\]
It is straightforward to transform a rank order into derived paired
comparison data. Suppose the order vector of a respondent on four items
is $(c,a,b,d)$, then we know that item $c$ is preferred to item $a$,
item $a$ is preferred to item $b$ and so on.

However, true paired comparison data and derived paired comparison data
from ranks differ in two ways:
\begin{enumerate}[(1)]
\item[(1)] In true paired comparison tasks, respondents might be
inconsistent in their preferences, producing an intransitive pattern
where the respondent is not choice consistent. In ranking tasks
inconsistent response patterns cannot occur.
\item[(2)] The mode of
presenting the items is different for the two tasks. In ranking data
all items are presented at once, while in a paired comparison task all
item pairs are presented in turn. Accordingly, different effects
concerning the order of the presentation of the items may occur.
\end{enumerate}

\section{Modeling ranked data}\label{se4}

\subsection{Modeling a single paired comparison}\label{se41}

The standard approach to modeling paired comparisons is the
Bradley--Terry (BT) model [\citet{brad52}]. We define the response in a
single paired comparison $(ij)$ to be $Y_{ij}$.  It is assumed that the
probability of an item $i$ being preferred to $j$ depends on the
nonnegative parameters $\pi_i$ and $\pi_j$ of the items $i$ and $j$,
defined as follows:
\[
\hspace{4.6pt}P\{Y_{ij}=1\vert \pi_i,\pi_j\}= \frac{\pi_i}{\pi_i+\pi_j}
\]
and
\[
P\{Y_{ij}=-1\vert
\pi_i,\pi_j\}=\frac{\pi_j}{\pi_i+\pi_j},
\]
where we later ensure that the $\pi_i$ sum to one for identifiability.

Thus,
\begin{eqnarray}\label{single}
P\{Y_{ij}=y_{ij}\vert\pi_i,\pi_j\}
&=&
\biggl(\frac{\pi_i}{\pi_i+\pi_j}\biggr)^{(1+y_{ij})/2}\biggl(\frac{\pi_j}{\pi_i+\pi_j}\biggr)^{(1-y_{ij})/2}\nonumber
\\[-8pt]\\[-8pt]
&=&
c_{ij}\biggl(\frac{\sqrt{\pi_i}}{\sqrt{\pi_j}}\biggr)^{y_{ij}},\nonumber
\end{eqnarray}
with $y_{ij}\in \{1,-1\}$ and with a constant
$c_{ij}^{-1}=\sqrt{\pi_i /\pi_j}+\sqrt{\pi_j /\pi_i}$
which does not depend on $y_{ij}$. We now reparameterize $\pi_i$ as
$\lambda_i=\frac{1}{2}\ln\pi_i$ or $\pi_i=\exp(2\lambda_i)$. Equation
(\ref{single}) then becomes
\begin{equation}\label{single2}
P\{Y_{ij}=y_{ij}\vert\lambda_i,\lambda_j\}=c_{ij}\exp\bigl(y_{ij}(\lambda_i-\lambda_j)\bigr)
\end{equation}
with $c_{ij}^{-1}=\exp(\lambda_i-\lambda_j)-\exp(\lambda_j-\lambda_i)$.

\subsection{Response patterns}\label{se42}

When transforming ranked data to paired comparison data with $J$ items,
we form all possible pairs of items. The number of such pairs
 is ${J \choose 2}$ and can be
ordered in a standard sequence:
$(12),(13),\dots,\break(1J); (23),(24),\dots,(2J);\ldots;((J-1)J)$. The
ranking outcome can therefore be recorded as a paired comparison
response pattern vector denoted by ${\bf
y}=(y_{12},y_{13},\dots,y_{J-1,J}) $ and consists of a series of $1$'s
and $-1$'s representing the values of the $y_{ij}$'s.

In the case of a true paired comparison task where all possible
comparisons are made, the number of all possible response patterns
 is given by the number of possible outcomes to the power of
the number of paired comparisons. If $y_{ij}$ can take only two values,
there are $2^{J \choose 2}$ possible response patterns in the space
$\Omega$. However, these response patterns also include intransitive
patterns which can not be generated from a ranking task. Removing these
intransitive patterns, the total number of patterns  is considerably
reduced to $ L= J!$. The space of transitive patterns is denoted by
$\Omega^T $.
For instance, the intransitive paired comparison pattern ($1 \succ 2$,
$2 \succ 3$, $3 \succ 1$) has no correspondence with any pattern
generated from ranking three items, since ranking patterns are
transitive by nature. Incorporation of intransitive patterns in the
contingency table would generate structural zeros and neglecting them
leads to biased estimates. Therefore, the use of a simple BT model,
which corresponds to a pattern model including intransitive patterns,
is not appropriate. Moreover, the dependence introduced by rankings
transformed to paired comparisons would not be addressed properly. For
instance, assuming independence, and in the simple case of three items,
given $Y_{12} = 1$, $Y_{23} = 1$, the probability of $Y_{13} = 1$ is
one, whereas the probability of $Y_{13} = -1$ is zero. However,
modeling the probabilities of whole response patterns
and reducing the number of possible patterns to those which are
transitive removes these dependencies.
We want to emphasize that we only consider complete rankings throughout
the paper. It is possible, however, to allow for partial rankings where
only a subset of items is ranked (see Section \ref{se8}).

\subsection{Modeling and estimation of transitive response patterns}\label{se43}

The probability for observing a sequence of paired comparisons ${\bf
y}$
is defined by
\[
P(\mathbf{y})= P(y_{12},y_{13},\ldots)=\prod_{i<j} P(y_{ij}),
\]
assuming independence between the comparisons. Using the probabilities
for a
single paired comparison defined in (\ref{single}), we then get
\begin{equation}\label{joint1}
P(\mathbf{y})=\prod_{i<j} c_{ij}\exp\bigl(y_{ij}(\lambda_i-\lambda_j)\bigr)
\end{equation}
or, equivalently,
\[
P(\mathbf{y})=\eta_{\mathbf{y}}\prod_{i<j} c_{ij}\qquad\mbox{with }\eta_{\mathbf{y}}=\exp\sum_{i<j}y_{ij}(\lambda_i-\lambda_j).
\]

Parameter estimation is based on multinomial sampling over the
transitive paired comparison patterns where it is supposed that each of
the $N$ respondents have completely ranked all $J$ items  and thus
contribute to one of the $L$ transitive response patterns. The
probability for observing a certain response pattern ${\mathbf{y}_{\ell}}$,
$\ell=1,\ldots,L$, given $J$ comparisons and transitive relations
only, is given as
\begin{equation}\label{mult.model}
\qquad P({\bf y_\ell}\vert J,\Omega^T)
=
\frac{P({\bf y}_\ell)}{\sum_{\ell'=1}^L P({\bf y}_{\ell'})}
=
\frac{\exp(\eta_\ell)\prod_{i<j}c_{ij}}{\sum_{\ell'}\exp(\eta_{\ell'}) \prod_{i<j}c_{ij}}
=
\frac{\exp(\eta_\ell)}{\sum_{\ell'}\exp(\eta_{\ell'})},
\end{equation}
where
\begin{equation}\label{eta0}
\eta_{\ell}=\sum_{i<j}y_{ij;\ell}(\lambda_i-\lambda_j).
\end{equation}
To ease notation, $P({\mathbf{y}_\ell} \vert J, \Omega^T)$ is denoted as
$P({\mathbf{y}_\ell})$ throughout the paper.

Let $n_{\ell}$ be the number of times the response pattern $\ell$ is
observed, then the $n_{\ell}$'s are multinomially distributed where $N
= \sum_{\ell} n_{\ell}$ is the total number of respondents and the
probability $P({\mathbf{y}_\ell})$ for a certain response pattern $\ell$ is
given in (\ref{joint1}).

Thus, the likelihood function is
\[
\mathcal{L} = \prod_{\ell} P({\mathbf{y}_\ell})^{n_{\ell}}.
\]

The parameters $\lambda_j$  can be estimated (using suitable parameter
restrictions, e.g., setting the last parameter to zero for
identifiability) by using standard software such as the
\texttt{prefmod} package in \textsf{R} [\citet{hatz09}]. To fit the
model, a variable containing the counts $n_{\ell}$ and a specific
design matrix $\mathbf{X}$ both need to be set up. The method corresponds to
a Poisson log-linear formulation of model (\ref{mult.model}) which is
described in detail in \citet{ditt07}, who also describe the more
general case when undecided responses can occur.

All parameters in $\eta$ have interpretation in terms of log odds.
Comparing two response patterns $\ell$ and $\ell'$ where only one
$y_{ij}$ differs, that is, $y_{ij;\ell} =1$ and $y_{ij;\ell'} =-1$, the
log odds are $ \ln(P({\mathbf{y}_\ell}) / P({\mathbf{y}_{\ell'}})) = \eta_\ell -
\eta_\ell' = 2 (\lambda_i - \lambda_j)$. If the item $j$ is the
reference item $J$, the odds reduce to $\exp (2\lambda_i) $.

Estimates of the worths $ \hat{\pi}_{j}$ are calculated through the
expression
\[
 \pi_j = \frac{\exp({2\lambda_j})}{\sum_{j}\exp({2\lambda_j})}
\]
to ensure that the sum of the worths is equal to 1.

\subsection{Respondent covariates in ranked data}\label{se44}

In most practical applications it is important to determine if the
importance of items depend on respondent covariates. This can be viewed
as a mixture of experts model. \citet{gorm08b} give an example
analyzing ranked data using a choice-based modeling approach.
Initially, we consider categorical covariates only. In this case, each
distinct combination of covariates observed will form a covariate set;
assume that there are $K$ such sets ($1 < K \leq N$). For example, with
two factors AGE (with four levels) and SEX (with two levels), there
will be eight covariate sets. To model the effect of the covariates,
the $J! = L$ response patterns now become $LK$ response patterns. The
number of times the $\ell$th response pattern occurs within each
covariate set $k$ is
denoted by $n_{\ell k}$. 
The linear predictor $\eta$ becomes
\begin{equation}\label{fixed1}
\eta_{\ell k}=\sum_{i<j} y_{ij;\ell k}(\lambda_{ik}-\lambda_{jk}).
\end{equation}
Each $\lambda_{jk}$ is an interaction effect of the item $j$ and the
covariates. Thus, two covariates $A$ and $B$ could potentially lead to
the following effects $\lambda_{j.A}+\lambda_{j.B}+\lambda_{j.A.B}$ if
an interaction effect on the items between $A$ and $B$ needs to be
considered.
%
%

With continuous covariates, in general, each respondent will be likely
to  have his/her own distinct set of covariates, and $K$ will usually
be close to $N$. In the particular example of a single covariate $x$,
the linear predictor of the model generalizes to be of the form
\[
\eta_{\ell k}=\sum_{i<j} y_{ij;\ell k}(\lambda_{i}+x_k\beta_{i}-\lambda_{j}-x_k\beta_{j}).
\]

\section{The random effects model}\label{se5}

While the previous section has allowed for known covariates, there may
be other variables which are unmeasured or omitted from the data set,
and these will produce heterogeneity between respondents in the item
parameters. One common way to account for such heterogeneity is to
introduce random effects for each respondent. These random effects
would adjust each item parameter up or down to allow for these missing
covariates and, thus, we need $J$ random effect components,  one for
each of the items being ranked.

We now extend the above model to allow for random effects. As before,
we work  with data aggregated into patterns and covariate sets. For
each covariate set and response pattern we need to specify $J$ random
effect components $\delta_{jlk}$. The linear predictor now becomes
\begin{equation}\label{rand1}
\eta_{{\ell}k}=\sum_{i<j}y_{ij;{\ell}k}(\lambda_{ik}+\delta_{i{\ell}k}-\lambda_{jk}-\delta_{j{\ell}k}).
\end{equation}
On the worth scale, the random effects become multiplicative, which
will multiply the worths by adjustment factors, shifting the worth for
each item up or down in an unique way for each ${\ell}k $ combination.
We set  $\delta_{J{\ell}k}$  to be zero for identifiability, and we
define
\[
{\bolds\delta}_{{\ell}k}=(\delta_{1{\ell}k},\delta_{2{\ell}k},\ldots,\delta_{J-1;{\ell}k}),
\]
a $(J-1)$-component random effect vector for each combination of
response pattern and covariate pattern.

Integrating over the unknown $(J-1)$-component random effects, the
likelihood then becomes
\[
\mathcal{L}
=
\prod_{{\ell}k}\biggl(\int_{-\infty}^{\infty}\ldots\int_{-\infty}^{\infty}
P( y_{{\ell}k}|{\bolds\delta}_{{\ell}k})
g({\bolds\delta}_{{\ell}k})\,
d\delta_{1{\ell}k}\,
d\delta_{ 2{\ell}k}
\cdots\,
d\delta_{J-1;{\ell}k}\biggr)^{n_{{\ell}k}},
\]
where $ g({\bolds\delta}_{{\ell}k})$ is the multivariate probability
density function or mixing distribution of the random effects vector.
For dealing with the multivariate random effect, \citet{hartzel01}
suggest a number of possible approaches. The first approach is to
assume multivariate normality for $g(\cdot)$:
 ${\bolds\delta}_{{\ell}k} \sim {\operatorname{MVN}}(0,{\bolds\Sigma})$, where ${\bolds\Sigma}$ is an unknown $(J-1)\times(J-1)$
covariance matrix which would be estimated from the data. For example,
\citet{coull00} explored a multivariate binomial logit-normal
distribution, where the mixing distribution is multivariate normal.

An alternative method, and one which we explore in this paper, is to
adopt a nonparametric solution. This solution replaces the parametric
multivariate normal distribution by a series of mass point components
with unknown mass or probability, and unknown location. This
nonparametric maximum likelihood (NPML) technique [\citet{mallet86}; \citet{aitkin96}] has the advantage of being able to identify subpopulations
of the respondents with specific response patterns, as well as
identifying the effect of respondent covariates on these patterns.
  The mass-point approach is in fact a mixture model, with the earlier
multinomial covariate model being replaced by a mixture of
multinomials.

Initially, we suppose that the number of components is known and is set
to $R$. Then we have $R $ mass-point vectors; a typical mass point
component $r$ would have unknown mass-point locations
\[
{\bolds\delta}_{r}=(\delta_{1r}, \delta_{2r}, \dots, \delta_{J-1;r})
\]
and unknown component probability $q_{r}$. If $R$ is small, this
substantially simplifies  the problem by replacing a $ J-1$ dimensional
integral  with  a sum over $R$ terms.

The likelihood now becomes
\begin{equation}\label{lclikelihood}
\mathcal{L}=\prod_{{\ell}k}\Biggl(\sum_{r=1}^R q_{r}P_{\ell kr}({\mathbf{y}}_{{\ell}k}|{\bolds\delta_{r}})\Biggr)^{n_{{\ell}k}}\qquad\mbox{where }
\sum_\ell P_{\ell kr}=1,\ \forall k,r.
\end{equation}

 The model can be interpreted in two ways. If we consider the discrete
mass point components as approximating an underlying multivariate
distribution, then we should ignore any interpretation of the mixing
structure and interpret the $\lambda_{jk}$ alone. However, we can also
think of the model as representing underlying subpopulations (or latent
classes) of the respondents, and we can then interpret the
$\delta_{jr}$ (which for a specific latent class $r$ gives the extra
increase or decrease in item $j$'s parameter over the reference latent
class $R$).

We determine the number of mass point components by choosing the model
which minimizes the Bayesian Information Criterion (BIC) proposed by
\citet{schwartz78}, which provides a penalty on the deviance which is a
function of the number of pattern--covariate sets,
\[
\mathit{BIC}=-2\ln\mathcal{L}+p\ln(LK),
\]
where $LK$ represents the number of pattern--covariate combinations and
$p$ is the number of parameters in the model.

We need to make clear that the likelihood in (\ref{lclikelihood}) does
not necessarily account for the complex sampling design in the
Eurobarometer survey. As the latent classes account for heterogeneity,
it is likely that some of the latent classes will reflect clustering
and design effects. We return to this point later in the discussion
section.

\section{Algorithmic and computational issues}\label{sec:5}

The EM algorithm provides a computationally elegant solution to the
maximization of the the likelihood given in equation
(\ref{lclikelihood})
[\citet{aitkin96}]. The use of this algorithm is well known; we give
brief details here and  provide more detail in the online supplement
[\citet{francis10}]. We start by observing that we can view the problem
as a missing data problem, where the latent class membership indicators
for each pattern and covariate set are missing. We can write these as
$z_{{\ell}kr}$, with $z_{{\ell}kr} =1$ if pattern $\ell k$ belongs to
class $r$, and zero otherwise. The expected values of the $z$'s are
defined to be $w_{{\ell}kr}$ and are the posterior probabilities of
class membership for a respondent with pattern $\ell$ and covariate set
$k$. The E-step of the EM algorithm computes the conditional
expectation of the complete log-likelihood (involving the calculation
of the $w$'s), whereas the M-step maximizes the multinomial likelihood
with respect to the $\lambda$'s and $\delta$'s, given the current
expected values of the $z$'s, which can be carried out through an
expanded Poisson log-linear model with weights $w_{{\ell}kr}$. Fitting
the multinomial through a Poisson log-linear model necessitates that a
set of nuisance parameters be included in the linear predictor;  these
constrain the marginal totals for each covariate set to be equal to the
observed totals.

The $w_{{\ell}kr}$ can potentially be used to assign respondents to
classes. If a respondent belongs to covariate set $k$ and has response
pattern $\ell$, then we can assign to the class with the highest
posterior probability $w_{{\ell}kr}$ over the $r$ classes.

There are a number of specific  problems related to the fitting of
latent class models of this kind. The first is that of multiple maxima.
The EM algorithm guarantees convergence to a local maximum of the
likelihood, but not to a global solution.  To minimize this problem, we
chose fifty different sets of starting values for each value of $R$ and
for each covariate model, and quote the best value of  $-2\ln\mathcal{L} $ and BIC found.

The second problem relates to the well-known slow convergence of the EM
algorithm. A relatively tight convergence criterion of 0.001  on the
deviance difference was chosen to ensure convergence of parameter
estimates.

Additionally, the EM algorithm does not give correct standard errors
for the parameters, as the method assumes that the $z$'s are known
rather than estimated. Two solutions are used in this paper. First, it
is possible to adopt a hybrid scheme where the EM algorithm is used to
obtain convergence, and then a series of Gauss--Newton steps are used to
obtain the full Hessian matrix [\citet{aitkinaitkin96}].   A second
method which is appropriate where the likelihood is likely to be
nonquadratic is to use a procedure described by \citet{aitkin94} and
\citet{dietz95} to obtain correct standard errors. This sets the Wald
test statistic equal to the likelihood ratio chi-squared statistic
obtained by equating one of the parameters in the model to zero. From
the Wald-test statistic, the appropriate standard error is obtained for
the $\lambda$'s associated with effect $X$,
\[
\mathit{s.e.}(\hat{\lambda}_{jX})=\frac{\hat{\lambda}_{jX}}{\sqrt{2\ln L(\lambda_{jX}=\hat{\lambda}_{jX})-2\ln L(\lambda_{jX}=0)}} .
\]
It is important that this second procedure is carried out by using as
starting values the final estimates of $w_{{\ell}kr}$ obtained from the
final model. This will ensure that the algorithm will not converge to a
local maximum with higher deviance.

Both methods have advantages. The first method, while computationally
complex, gives asymptotic standard errors for all estimated parameters,
provided that good starting values are used for the Gauss--Newton steps.
The second method has the advantage of providing a standard error which
gives a $t$-test $p$-value equivalent to the appropriate likelihood ratio
test. However, label switching problems can occur in using the second
method especially when setting, for example, a specific delta parameter
to zero.

Finally, for large $K$, the algorithm will take longer to converge and
require more memory, both because of the need to increase the size of
the table [${\mathbf{y}}_{{\ell}k}$] to be analyzed, and the large number of
lambda parameters $\lambda_{jk}$ and nuisance parameters needed to fit
the multinomial by means of a Poisson log-linear model. Numerical
procedures such as those described in \citet{hatzfran04} can be used to
remove the need to estimate the nuisance parameters and to speed
convergence.

For this paper, models were fitted using  the \texttt{pattnpml.fit}
function of the \textsf{R} [\citet{Rd2008}] package \texttt{prefmod}
[\citet{hatz09}]. The \texttt{pattnpml.fit} function is a modification
of the \texttt{alldist} function in the package \texttt{npmlreg}
[\citet{einbeck07}], and has been adapted to allow multiple random
effects terms and more flexibility in the choice of start values.

\section{Data analysis}\label{se7}

We now apply the above model to the Eurobarometer question. There are
$12216$ complete responses in the data set. We choose covariates of
\texttt{AGE} (4 levels: 15--24, 25--39, 40--54 and 55$+$) and \texttt{SEX}
(2 levels: male, female) to illustrate the methodology. There are other
important covariates, such as educational level, income and country of
origin, which have been identified by \citet{christ01}, but we exclude
these in this illustration to ensure that omitted variables and random
effects are needed in the analysis. Of the $720$ response patterns, the
most popular response is $(\mathit{TV}, \mathit{Rad}, \mathit{Press}, \mathit{SciM}, \mathit{WWW}, \mathit{Edu})$
with 526 respondents, followed by $(\mathit{TV}, \mathit{Rad}, \mathit{Press}, \mathit{SciM}, \mathit{Edu}, \mathit{WWW})$ with 507. Only 70 (9.7\%) of the response patterns are not used
at all by the respondents.

\subsection{Modeling ``Sources of science information'' data}\label{se71}

Our model fitting strategy was to determine a covariate model using
simple fixed effects models (that is, without random effects terms),
then fixing the covariates in the model and increasing the number of
mass point vectors to allow for the unknown random effects distribution
to be approximated by the nonparametric mass point components. We
started with the ``null'' model without covariates (\ref{eta0}), which
estimated a common set of item parameters for all respondents. We then
included the respondent covariates \texttt{AGE} and \texttt{SEX} and
examined possible main effect and interaction models. Equation
(\ref{fixed1}) reminds us that when we refer to the model \texttt{SEX},
we are in fact fitting an interaction term between the \textit{items}
$(\mathit{TV}, \mathit{Rad}, \mathit{Press}, \mathit{SciM}, \mathit{WWW}, \mathit{Edu})$ and \texttt{SEX} and
specifying 12 interaction parameters in the model: $\lambda_{\mathit{TV.SEX}}$,
$\lambda_{\mathit{Rad.SEX}}$, $\lambda_{\mathit{Press.SEX}}$, $\lambda_{\mathit{SciM.SEX}}$,
$\lambda_{\mathit{WWW.SEX}}$ and $\lambda_{\mathit{Edu.SEX}}$. Two of these parameters
($\lambda_{\mathit{Edu.male}}$ and $\lambda_{\mathit{Edu.female}}$) are constrained to
zero for identifiability. We examined changes in deviance and the
Bayesian information criterion BIC [\citet{schwartz78}] to compare model
fits and to find the best model (that is, the model with the lowest
BIC). To allow deviances and BIC values to be compared, we fitted
models to the same sized table [${\mathbf{y}}_{{\ell}k}$]---with eight
covariate sets, all model fits included eight nuisance parameters (the
\texttt{AGE} by \texttt{SEX} interaction).

\begin{table}[b]
\caption{Fixed effect models}\label{new1}
\begin{tabular}{@{}lccc@{}}
\hline
\textbf{Model}&\textbf{Deviance}&\textbf{No. of parameters}&\textbf{BIC}\\
\hline
Null&21,293&13&21,406\\
\texttt{AGE}&18,078&28&18,321\\
\texttt{SEX}&21,041&18&21,197\\
\texttt{AGE$+$SEX}&17,815&33&18,100\\
\texttt{AGE$+$SEX$+$AGE:SEX}&17,790&48&18,206\\
\hline
\end{tabular}
\end{table}

As can be seen in Table \ref{new1}, the main effects model
\texttt{AGE$+$SEX} has the lowest  BIC ($=18\mbox{,}100$) and there is no need
for the interaction between \texttt{AGE} and \texttt{SEX}. In the
paired comparison model this means both factors \texttt{AGE} and
\texttt{SEX }have a separate effect on the item parameters and,
therefore, the worths of the items change with \texttt{AGE} and
\texttt{SEX}.

We can consider two forms of random effects models. We first
investigated whether a simple random effects model without covariates
provides a better explanation than the fixed effects model. The model
without covariates is equivalent to fitting a latent class model to the
data. We then fitted random effects models with fixed covariate terms
\texttt{AGE$+$SEX}, and tested whether the covariates are still
important.

\begin{table}
\caption{NPML random effects models with and without covariates}
\label{new2}
\begin{tabular}{@{}lccc@{}c@{}cccc@{}}
\hline
\multicolumn{4}{@{}c@{}}{\textbf{(a) Without covariates}}&&\multicolumn{4}{@{}c@{}}{\textbf{(b) With AGE and SEX as covariates}}\\[-5pt]
\multicolumn{4}{@{}c@{}}{\hrulefill}&&\multicolumn{4}{@{}c@{}}{\hrulefill}\\
\textbf{No. of}&&\textbf{No. of}&&&&\textbf{No. of}&&\\
\textbf{mass}&&\textbf{para-}&&&&\textbf{para-}&&\textbf{Final}\\
\textbf{points $\bolds{r}$}&\textbf{Deviance}&\textbf{meters}&\textbf{BIC}&&\textbf{Deviance}&\textbf{meters}&\textbf{BIC}&\textbf{model}\\
\hline
1&21,293   & 13   &    21,406 &\hspace*{6pt}& 17,815   &33 &  18,100   &   \\
2&12,494   & 18   &    12,650             && 10,731   &38 &  11,060   &   \\
3&10,252   & 23   &    10,451             &&  \phantom{0,}9056    &43 &   \phantom{0,}9428    &   \\
4& \phantom{0,}9792    & 28   &    10,035             &&  \phantom{0,}8836    &48 &   \phantom{0,}9252    &   \\
5& \phantom{0,}9544    & 33   &     \phantom{0,}9830  &&  \phantom{0,}8729    &53 &   \phantom{0,}9187    &   \\
6& \phantom{0,}9387    & 38   &     \phantom{0,}9716  &&  \phantom{0,}8667    &58 &   \phantom{0,}9170    & \maltese \\
7& \phantom{0,}9302    & 43   &     \phantom{0,}9674  &&  \phantom{0,}8636    &63 &   \phantom{0,}9182    &   \\
8& \phantom{0,}9277    & 48   &     \phantom{0,}9693  &&  \phantom{0,}8623    &68 &   \phantom{0,}9212    &   \\
\hline
\end{tabular}
\end{table}

The model with a single mass point component means that all respondents
are in one latent class, and corresponds to the null fixed effect model
(deviance $ = 21{,}293 $). Increasing the number of mass point components
(Table \ref{new2}a), we observed that the BIC steadily decreases with
no sign of a minimum being reached. We stopped at eight mass point
components, as we were not specifically interested in determining the
number needed for the model without covariates. However, we can observe
two features. First, through examination of BIC values, the latent
class model with two (BIC $= 12{,}650$) or more components fits
substantially better than the covariate model without random effects
\texttt{AGE$+$SEX} (BIC $= 18\mbox{,}100$). Second, a~large number of latent
classes will be needed to fully represent omitted covariates (which in
this model also include \texttt{AGE} and \texttt{SEX}).

Can a mixed model provide a way forward, and are the measured
covariates still important given the importance of latent class
structure? Table  \ref{new2}b shows the results obtained by fitting the
random effects model with fixed covariates \texttt{AGE$+$SEX}. With one
mass point component, the model corresponds to the fixed effects
\texttt{AGE$+$SEX} model in Table \ref{new1}. The minimum BIC is found at
$r=6$ classes; the deviance is substantially less than the deviance for
$r=8$ classes with no covariates. It appears that the fixed effects
provide additional explanatory power, and this becomes our final model.
Removal of \texttt{AGE} and \texttt{SEX} in turn produces a large
significant change in deviance and the covariate model cannot be
simplified.

\begin{table}
\tablewidth=265pt
\caption{Parameter estimates for $\lambda_{\mathit{SciM.AGE}}$ for fixed and random
effects models: \texttt{AGE$+$SEX}}\label{agepes}
\begin{tabular}{@{}lccccc@{}}
\hline
&\multicolumn{2}{c}{\textbf{(a) Fixed effects model}}&\multicolumn{3}{c@{}}{\textbf{(b) Mixture random effects model}}\\[-5pt]
&\multicolumn{2}{c}{\hrulefill}&\multicolumn{3}{c@{}}{\hrulefill}\\
&&&&\textbf{Raw EM}&\textbf{Corrected}\\
&&\textbf{Standard}&&\textbf{standard}&\textbf{standard}\\
\textbf{AGE}&\textbf{Estimate}&\textbf{error}&\textbf{Estimate}&\textbf{error}&\textbf{error}\\
\hline
15--24&0\phantom{.000}&---&0\phantom{.000}&---&---\\
25--39&0.165&0.011&0.169&0.012&0.018\\
40--54&0.201&0.012&0.198&0.013&0.019\\
55$+$&0.219&0.011&0.208&0.013&0.019\\
\hline
\end{tabular}
\end{table}

We can interpret the final fitted model in two ways. We can treat the
mass point components as approximating an unknown multivariate
distribution, and focus attention primarily on the covariates. As an
illustration, Table \ref{agepes} shows the estimates for
$\lambda_{\mathit{SciM.AGE}}$ for both the fixed effects model and the final
random effects model, with a reference category of school/university
($\mathit{Edu}$). We can see that as age increases, the preference for scientific
magazines compared to school/university as a source of information
increases---this is true for both fixed and random effects models, but
the effects are attenuated for the random effects model. Other age
parameters (not shown) show a relative preference decrease in the use
of the internet ($\mathit{WWW}$), and an increase in TV, newspapers ($\mathit{Press}$) and
scientific magazines compared with school/university. Unadjusted and
corrected standard errors [\citet{aitkin94}] are given for the random
effects model and we can observe that the uncorrected  and corrected
standard errors are relatively close in this example.

From the estimates of $\lambda_{\mathit{items.SEX}}$ (not shown), we can also
conclude that the preference for both scientific magazines and the
internet relative to school/univer\-sity  is significantly lower for
females than for males.

\begin{table}[b]
\caption{Proportions in the six classes} \label{classprop}
\begin{tabular}{@{}lcccccc@{}}
\hline
&\textbf{Class 1}&\textbf{Class 2}&\textbf{Class 3}&\textbf{Class 4}&\textbf{Class 5}&\textbf{Class 6}\\
\hline
Proportions of patterns&0.3156&0.1289&0.3329&0.0583&0.0984&0.0659\\
Proportions of respondents&0.1808&0.0739&0.2460&0.0716&0.1407&0.2890\\
\hline
\end{tabular}
\end{table}

It is also possible to proceed by treating the mass point components as
latent classes. Table \ref{classprop} shows the estimated proportions
of patterns $\hat{q}_r$ (which are obtained directly from the
algorithm) and the estimated proportions of respondents which are
weighted averages of the posterior probabilities  of pattern membership
in each class ($w_{{\ell}kr}$), weighted by the proportion of
respondents in each pattern. Equations (3) and (4)  in the online
supplement provide further details. Examining the proportions of
respondents, we see that class 6 is the largest class with just under
29\% of respondents, followed by class 3 with about 25\% and class 1
with just over 18\%.

\begin{figure}

\includegraphics{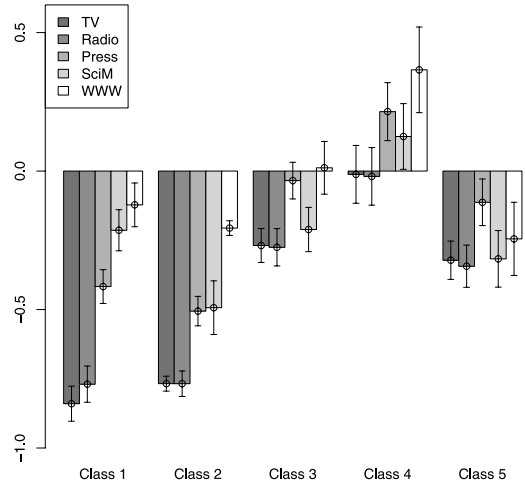}

\caption{Parameter estimates for $\bolds{\delta}_r$ and
95\% confidence intervals based on corrected standard errors.}\label{deltaplot} 
\end{figure}

Figure \ref{deltaplot} shows the estimated random effect components
$\bolds{\delta}_r$ for all items and all classes  (apart for the reference
item $J$ and class $R$ which are set to zero) including 95\% confidence
intervals based on the corrected estimated standard errors.
The bars $(\delta_{jr})$ are half the log odds ratios 
comparing the extra effect of item $j$  to the reference item $J$
(education) and for class $r$ related to the reference class $R$ (class~6).

It can be seen, for example, that for class 1  the odds for TV  and
Radio are substantially lower than for Education compared to class 6
[TV: $\exp (-0.84 \cdot 2) = 0.186 $, Radio: $\exp (-0.77 \cdot 2) =
0.215 $]. In class 4 the odds for Press compared to Education are about
$1.5$ times higher and  for WWW 2.1 times higher than in class 6
[Press: $\exp ( 0.21 \cdot 2)$, WWW: $\exp ( 0.37 \cdot 2)$].

\begin{figure}

\includegraphics{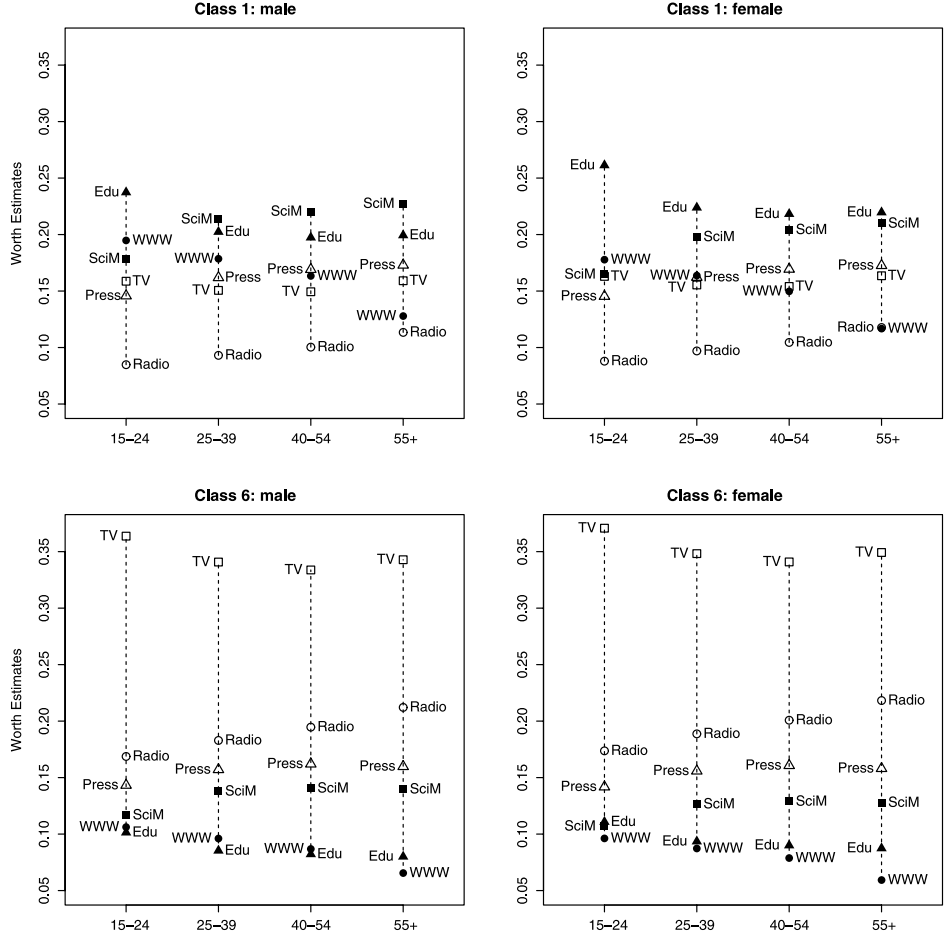}

\caption{Item worths by age and gender for two
extreme latent classes.}\label{worthplots}
\end{figure}

Figure \ref{worthplots} shows, for males and for females, the plotted
worths against age for each of the six sources of information, for two
of the six latent classes. We see that the two classes represent
different preference patterns in the data. Class 6 represents a large
subpopulation who prefer to obtain most of their scientific information
from nontext and nonscholarly sources.  For all age groups and for both
males and females, TV is the most preferred source, with radio the
second most preferred and increasing in preference with age.  Class 1,
in contrast, represents a smaller subpopulation which prefers academic
sources of information over more popular information sources. In this
class, for all but the youngest age group, scientific magazines and
school/university sources rank in the top two places (with scientific
magazines winning out over school/university for males but not for
females). For the youngest age group, the school/university followed by
the internet are preferred for both males and females. Class 3, the
second largest group (not shown), shows a latent class which is similar
to class 6 but with a different second preference. TV is still the most
preferred source, followed by newspapers and the radio for the three
older age groups. For the youngest age group, radio declines in
preference and the third preferred source becomes the internet for
males and school/university for females. In terms of the other classes
which are not displayed, classes 4 and 5 also have TV in first place,
but with different orderings of other sources in other places. Class 2
(7\%) prefers school/university as the most preferred source of
information but with TV in second place.

\subsection{Analysis of class membership}\label{se72}

It is to be expected that relevant variables not included in the model
are absorbed in the latent classes. This relates to variables which are
(i) known but for various reasons not accounted for (e.g., variables
with many categories making computation unfeasible or impossible) and
also to (ii) possibly unknown sources of variation. In the
Eurobarometer survey, for example, there is a complex five-level
clustering design of households within address clusters within PSUs
within urbanization and administrative region strata and within
countries. While some of these variables are present in the data set,
others are not. In addition, each country has used a different coding
scheme for determining degree of urbanization. This means that a full
multilevel analysis taking account of all design components is not
possible. However, it could be argued that the most important strata
are degree of urbanization and country, and these two levels would
account for most variability within the clustered sample. We therefore
examine the effect of these two variables below.

To evaluate the effect of known variables, a post-hoc analysis may be
performed by analyzing their association with the respondents' class
memberships. Two approaches are possible which use different
definitions of class membership. We illustrate using two covariates not
in the model but which are used in the sample design---degree of
urbanization and country. For degree of urbanization, we adopt a common
three-level categorization  which is consistent across countries. We
use 15 countries rather than 17 for this investigation, combining East
and West Germany (D), and Great Britain and Northern Ireland (GB). The
remaining countries are labeled  by their international licence plate
country code.

\begin{figure}[b]

\includegraphics{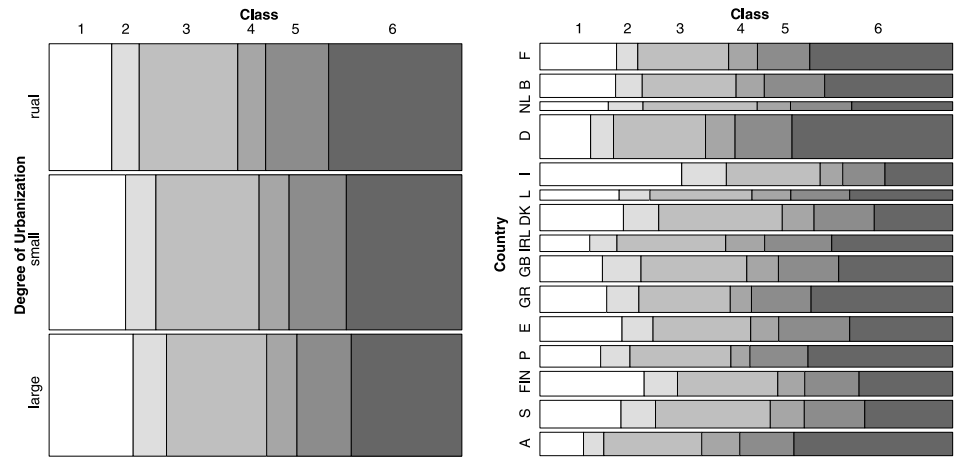}

\caption{Mosaic plots showing expected class membership and degree of urbanization
(left) and country (right).}\label{mosaic1}
\end{figure}

The first method uses the posterior probabilities of class memberships
to construct the expected number of respondents in each class within
each category of the covariate of interest [see equation (4) in the
online supplement]. We present two mosaic plots
[\citet{hartigan1984mosaic}] which cross-classify the expected class
membership with degree of urbanization and with country (displayed in
Figure \ref{mosaic1}).

In examining the degree of urbanization mosaic plot, it can be seen
that the proportion of rural residents are underrepresented in class
$1$ and have a higher proportion in class $6$ as opposed to residents
of large cities.
The country mosaic plot shows much greater variability. Respondents in
Italy, for example, are far less likely to belong to latent class $6$
and far more likely to belong to latent class $1$. In contrast,
respondents in Austria and Germany are far more likely to belong to
class $6$. One explanation for this variability might be the varying
quality of TV across countries in broadcasting science information,
coupled with a large number of excellent science magazines in Italy.

A second approach, as mentioned in Section \ref{sec:5}, assigns the
respondents (who belong to covariate set $k$ and have response pattern
$\ell$) directly to the class with the highest posterior probability
$\max_r(w_{{\ell}kr})$. Following this procedure, we can obtain a
response variable with categories according to the classes and
investigate the effects of some variables not included in the model via
a multinomial regression model. We then form a cross-classified table
of assigned class by country and by degree of urbanization to evaluate
possible influences due to part of the multistage sampling design. By
fitting a multinomial model, we found a strong interaction effect
between degree of urbanization and
country. 

\begin{figure}

\includegraphics{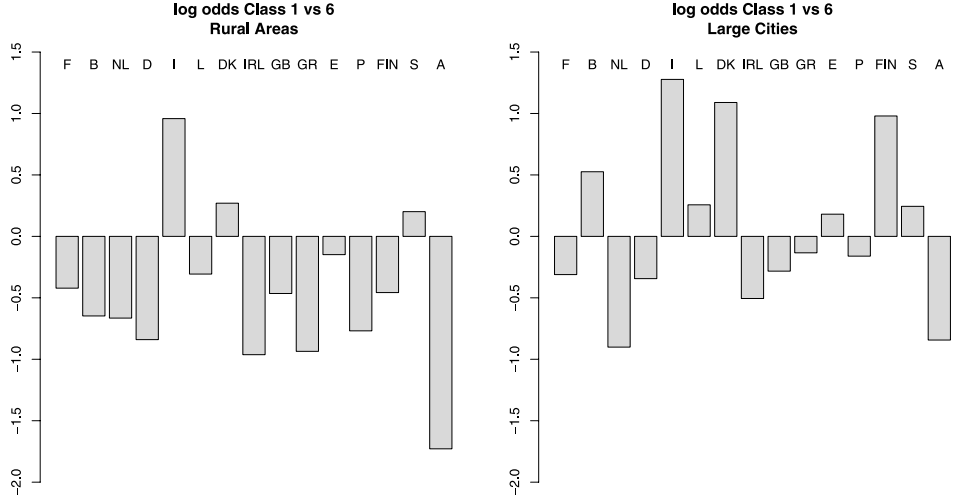}

\caption{Plot of observed log-odds ratios for class 1 against class 6 for assigned
class membership
classified by country and degree of urbanization.}\label{logodds1}
\end{figure}

This interaction can be visualized by examining observed log-odds
ratios in the constructed table. Figure \ref{logodds1} shows the
observed log-odds ratios comparing classes $1$--$6$ for the 15
countries both for rural areas and for large cities. We can notice, for
example, that Italy has a positive log-odds ratio for both rural areas
and large cities, indicating the relative underrepresentation of class
6 is true both for urban and rural locations. In other countries such
as Finland, class 6 is more prevalent in rural areas, and class 1 in
large cities.

\section{Discussion}\label{se8}
Random effects models are often necessary in models for ranked and
paired comparison data but the multivariate nature of random effects in
these type of models adds complexity. NPML methods of the type
described here provide a suitable way forward. The models give greater
insight into the nature of subgroups in the data set, but
interpretation  can be problematic because of the number of parameters
being estimated. We recommend the use of graphical displays on the
worth scale.

Diagnostic checks are important for these models. It is important to
examine the solution to check both that there are no overly small
latent classes, and also that the parameter estimates for each mass
point component are sufficiently separate [\citet{mclach99}]. Posterior
probabilities of component membership could also be examined in
relation to other covariates not in the model to aid interpretation of
the latent classes [\citet{kamak91}].

The basic model described in this paper can be extended in various
ways:
\begin{itemize}
\item Extensions to models which allow varying coefficients with latent
classes is straightforward. This model will allow for different
respondent covariate effects within each latent class. These random
coefficient models can be fitted by allowing interactions between the
latent class group and the covariates, but with the disadvantage of a
sizeable increase in the number of model parameters.
\item It is possible to extend the model to allow for tied ranks.  Such
data will lead to an underlying ordinal paired comparison model
[\citet{ditt04}]. \item Item covariates could also be included along
the lines suggested by \citet{dittrich98a}. \item The model presented
here needs to be extended to allow explicitly for more complex sampling
designs and other multilevel structures which may be present in the
data. Further research is needed on this topic. \item Finally,
incomplete or partial rankings could also be taken account of. This
would lead to a paired comparison model which allows for missing
comparisons within a response. The basic idea here is to extend the set
of response patterns to include patterns where certain comparisons are
not available. For partial rankings a composite link approach to this
problem has been described in \citet{dabicshatzs2009}; the general case
for paired comparisons with missing data is treated in \citet{ditt10}.
Unfortunately, the number of response patterns may increase
dramatically and, thus, this approach is computationally feasible only
for a small number of items.
\end{itemize}
In conclusion, our approach provides a methodology which  allows the
modeling of ranked data in many applied areas, allowing covariates to
be taken into account and latent classes to be detected. The underlying
paired comparison approach provides an attractive alternative to the
choice based models dominant in the literature.

\section*{Acknowledgments}\label{se9}

This research was supported by the ESRC  under the National Centre for
Research Methods initiative (Grant numbers RES-576-25-5020 and
RES-576-25-0019).  We would like to thank Walter Katzenbeisser for
helpful statistical discussions, and the referees and editors for
insightful suggestions. Eurobarometer questions are reproduced with the
license granted by its author, the European Commission,
Directorate-General for Information, Communication, Culture and
Audiovisual Media, 200 rue da le Loi, B-1049 Brussels, and by
permission of its publishers, the Office for Official Publications of
the European Communities, 2 rue Mercier, L-2985 Luxembourg (\copyright\,European Communities). The data set was provided by Zentralarchiv f\"ur
Empirische Sozialforschung, K\"oln (ZA 3509).

\begin{supplement} [id=suppA]
\stitle{The EM algorithm for NPML random effects in ranked data\break}
\slink[doi]{10.1214/10-AOAS366SUPP}
\slink[url]{http://lib.stat.cmu.edu/aoas/366/supplement.pdf}
\sdatatype{.pdf}
\sdescription{We provide a detailed description of
the use of the EM algorithm for fitting nonparametric random
effects for ranked data by maximum likelihood.}
\end{supplement}

\printaddresses

\end{document}